\documentclass[letterpaper]{article}
\usepackage{latexsym,amsfonts,amsmath,amsthm,amssymb}
\usepackage{pstricks}

\begin{document}

\title{Quantum-like Probabilistic Models outside Physics}
\author{Andrei Khrennikov\\
International Center for
Mathematical Modeling \\ in Physics and Cognitive Sciences\\
University of V\"axj\"o, S-35195, Sweden}

\maketitle
\begin{abstract}
We present a  quantum-like (QL) model in that
contexts (complexes of e.g. mental, social, biological, economic or even
political conditions) are represented by complex probability amplitudes.
This approach gives the possibility to apply the mathematical quantum formalism to
probabilities induced in any domain of science. In our model
quantum  randomness appears not as irreducible randomness (as it is
commonly accepted in conventional quantum mechanics, e.g., by von Neumann and Dirac), 
but as a consequence of obtaining incomplete information about a system. We pay main
attention to the QL description of processing of
incomplete information. Our QL  model can be useful in cognitive, social and political
sciences as well as economics and artificial intelligence. In this paper we consider in 
a more detail one special application -- QL  modeling of brain's
functioning. The brain is modeled as a QL-computer. 
\end{abstract}


\section{Introduction: Quantum Mechanics as Operation with Incomplete 
Information}

Let us assume that, in spite of a rather common opinion,  
quantum mechanics is not a complete theory. Thus the wave function does not provide 
a complete description of the state of a physical system. Hence we assume that 
the viewpoint of Einstein, De Broglie, Schr\"odinger, Bohm, Bell, Lamb, Lande, 
`t Hooft and other believers in the possibility to provide a more detailed description of quantum 
phenomena is correct, but the viewpoint of Bohr, Heisenberg, Pauli, Fock, Landau and other 
believers in the completeness of quantum mechanics (and impossibility to go beyond it) is wrong. 
We remark 
that the discussion on completeness/incompleteness of quantum mechanics is
 also known as the discussion about 
{\it hidden variables} -- additional parameters which are not encoded in the wave function.
In this paper we would not like to be involved into this great discussion, see e.g. 
\cite{Khrennikov/P02}, 
\cite{Khrennikov/P05} for recent debates. We proceed in a very pragmatic way by taking advantages
of the incompleteness
viewpoint on quantum mechanics. Thus we would not like to be waiting for until the end of the
Einstein-Bohr 
debate. This debate may take a few hundred years more.\footnote{Even if quantum mechanics 
is really incomplete it may be that hidden parameters would be found only on scales 
of time and space that would be approached 
not very soon, cf. \cite{Khrennikov/T}}

\medskip

{\it What are advantages of  
the Einstein's interpretation of the QM-formalism? }

\medskip

The essence of this formalism is not the description
of a special class of physical systems, so called quantum systems, having rather exotic and even 
mystical properties, but the possibility to operate with {\it incomplete
information} about systems. Thus according to Einstein one may apply the
formalism of quantum mechanics in any situation 
which can be characterized by incomplete description of systems. This (mathematical) formalism
could be used in any domain of science, cf. \cite{AR}--\cite{Khrennikov/book:2004}, 
\cite{Choustova}, \cite{Khrennikov/QLBrain},
: in cognitive and 
social sciences, economy, information
theory.  Systems under consideration need not be exotic. However, the complete description of them 
should be not available or ignored (by some reasons).

We repeat again that in this paper it is not claimed that quantum mechanics
is really incomplete as a physical 
theory. It is not the problem under consideration. We shall be totally satisfied 
by presenting an information  QL model such that 
it will be possible to interpret the wave function as an 
incomplete description of a system.
It might occur that our model could not be applied to quantum mechanics as a physical theory. 
However, we shall see that our model can be used {\it at least outside physics.} Therefore
we shall speak about a quantum-like (QL) and not quantum model.

We shall use Einstein's interpretation of the formalism of quantum mechanics. 
This is a special mathematical 
formalism for statistical decision making in the absence of complete information about systems. 
By using this formalism one permanently 
ignore huge amount of information. However, such an information processing does not induce 
chaos. It is extremely {\it consistent.} Thus the QL information cut off is done in a clever 
way. This is the main advantage of the QL processing of information. 

We also remark that (may be unfortunately) the mathematical formalism for operating with 
probabilistic data represented by complex amplitudes was originally discovered 
in a rather special domain of physics, so called quantum physics. This historical fact 
is the main barrier on the way of applications of  QL methods in other domains of science, e.g.
biology, psychology, sociology, economics. If one wants to apply somewhere the mathematical 
methods  of quantum mechanics (as e.g. Hameroff \cite{Hameroff1}, \cite{Hameroff2} 
and Penrose \cite{Penrose1}, \cite{Penrose2} 
did in study of brain's  
functioning) he would be typically 
constrained by the conventional interpretation of this formalism, namely, the 
orthodox Copenhagen interpretation. By this interpretation quantum mechanics is complete. 
There are no additional (hidden or ignored) parameters completing the wave function description.
Quantum randomness is not reducible to classical ensemble randomness, see 
von Neumann \cite{Neumann},  
Birkhoff and von Neumann \cite{Birkhoff/vonNeumann}.\footnote{The founders of the orthodox Copenhagen interpretation, 
Bohr and Heisenberg as well as e.g. Pauli and Landau, did not discuss so much probabilistic
aspects of this interpretation. This is an astonishing historical fact, since quantum mechanics
is a fundamentally probabilistic theory.}   
It is not easy to proceed with this interpretation to macroscopic applications.

In this paper we present a  contextualist statistical realistic model  for
QL representation of incomplete information for any kind of systems: 
biological, cognitive, social, political, economical. Then we concentrate our considerations
to cognitive  science and psychology
\cite{Khrennikov/book:2004}, \cite{Khrennikov/Supp}. In particular, we 
shall  describe cognitive experiments to check the QL structure of mental processes. 

The crucial role
is played by the {\it interference of probabilities for mental
observables.} Recently one such experiment based on recognition of
images was performed, see \cite{Conte}, \cite{Khrennikov/book:2004}. This experiment 
confirmed our prediction
on the QL behavior of mind. In our approach ``quantumness
of mind'' has no direct relation to the fact that the brain (as any
physical body) is composed of quantum particles, cf. \cite{Hameroff1}--\cite{Penrose2}.  
We invented a new terminology ``quantum-like (QL) mind.'' 

Cognitive QL-behavior is
characterized by a nonzero {\it coefficient of interference} $\lambda$
(``coefficient of supplementarity'').
This coefficient can be  found on the basis of statistical data.
There are predicted not only $\cos \theta$-interference of
probabilities, but also hyperbolic $\cosh \theta$-interference. The
latter interference  was never observed for physical systems, but we
could not exclude this possibility for
 cognitive systems. We propose a model of brain functioning as  a QL-computer.
 We shall  discuss the difference between quantum and QL-computers.

 From the very beginning we emphasize that our approach
has nothing to do with {\it quantum reductionism.} Of course, we do not
claim that our approach implies that quantum physical reduction of
mind is totally impossible. But our approach could explain the main
QL-feature of mind -- {\it interference of minds} --
without reduction of mental processes to quantum physical processes.
Regarding the quantum logic approach we can say that our contextual
statistical model is  close mathematically to some models of quantum
logic \cite{Mackey}, but interpretations of
mathematical formalisms are quite different.  The crucial point is
that in our probabilistic model it is possible to combine {\it
realism} with the main distinguishing features of quantum
probabilistic formalism such as {\it interference of probabilities,
Born's rule, complex probabilistic amplitudes, Hilbert state space,
and representation of (realistic) observables by operators.}

In the first version of this paper I did not present the references on the 
 original source of firefly in the box example. Andrei Grib told me about this example,
 but I had impression that this was a kind of quantum logic folklor. Recently
 Bob Coecke and Jaroslaw Pykacz told me the real story of this example.
 I am happy to complete my paper with corresponding references.
 
 It was proposed by Foulis who wanted to show that a macroscopic system, firefly,
can exhbit a QL-behavior which can be naturally represented in terms of quantum logics.
 First time this example
was published in Cohen's book \cite{Cohen}, a detalied presentation can be found in Foulis' 
paper \cite{Foulis}, 
see also Svozil \cite{Svozil1}. Later ``firefly in the box''
was generalized to a so called  generalized urn's model, by Wright \cite{ Wright} (psychologist).

\section{Levels of Organization of Matter and Information and their Formal Representations}

This issue is devoted to the principles and mechanisms which let matter to build 
structures at different levels, and their formal representations. We would like to extend 
this issue by considering information as a more general structure than matter, 
cf. \cite{KhrennikovAN}.
Thus material structures are just information structures of  a special type. 
In our approach it is more natural to speak about {\it principles and mechanisms 
which let information to build  structures at different levels, and their formal 
representations.} From this point of view quantum mechanics is simply a special formalism 
for operation in a consistent way with incomplete information about a system. Here a system 
need not be 
a material one. It can be a social or political system, a cognitive system, a system of economic 
or financial relations. 

The presence of OBSERVER collecting information about systems is always assumed 
in our QL model. Such an observer 
can be of any kind: cognitive or not, biological or mechanical. The essence of our approach is
that such an observer is able to obtain some information about a system under observation.
As was emphasized, in general this information is not complete. An observer may collect incomplete
information not only because it is really impossible to obtain complete information. It may occur 
that it would be convenient for an observer or a class of observers 
to ignore a part of information, e.g., about social or political processes.\footnote{We mention
that according to Freud's psychoanalysis human brain can even repress some ideas, so called hidden 
forbidden wishes and desires, and send them into the unconsciousness.}

Any system of observers with internal or external structure of self-organization should operate 
even with incomplete   information in a consistent way. The QL formalism provides such a 
possibility.

We speculate that observers might evolutionary develop the ability to operate with incomplete information
in the QL way --  for example, brains. In the latter case we should consider even 
{\it self-observations:} each brain performs measurements on itself. We formulate
the hypothesis on the QL structure of processing of mental information. Consequently,
the ability of QL operation might be developed by social systems and hence in economics and 
finances.

If we consider {\it Universe as an evolving information system,} then we shall evidently see a 
possibility that Universe might evolve in such a way that different levels of its organization
are coupled in the QL setting. In particular, in this model quantum physical reality is simply a 
level of organization (of information structures, in particular, particles and fields)
which is based on a consistent ignorance of information (signals, 
interactions) from a {\it prequantum level.} In its turn the latter may also play the role of the 
quantum level for a pre-prequantum one and so on. We obtain the model 
of Universe having many (infinitely many) levels of organization which are based on 
QL representations of information coming from the previous levels. 

In same way the brain might have  a few levels of transitions from the classical-like (CL) to 
the QL description.

At each level of representation of information  so called physical, biological, mental, 
social or economic laws are obtained only through partial ignorance of information coming from 
previous levels. In particular, in our model the quantum dynamical equation, the Schr\"odinger's
equation, can be obtained as a special (incomplete) representation of dynamics of classical 
probabilities, see section 12.

\section{V\"axj\"o Model: Probabilistic Model for Results of Observations}

 A general statistical realistic model for observables based on the contextual
viewpoint to probability will be presented. It will be shown that
classical as well as quantum probabilistic models can be obtained as
particular cases of our general contextual model, the {\it{V\"axj\"o
model}}.
    
This model is not reduced to the conventional, classical and quantum
    models. In particular, it contains a new statistical model: a model with
hyperbolic $cosh$-interference that induces  "hyperbolic quantum
mechanics" \cite{Khrennikov/book:2004}.

A physical, biological, social,  mental, genetic, economic, or financial
{\it context}  $C$ is  a complex of corresponding conditions.
Contexts are fundamental elements of any contextual statistical model. Thus construction  
of any model
$M$ should be started with fixing the collection of  contexts of this model.
Denote the collection of contexts
by the symbol ${\cal C}$ (so the family of contexts  ${\cal C}$ is determined by the 
model $M$ under consideration). In the mathematical formalism ${\cal C}$ is an abstract set
(of ``labels'' of contexts).  

We remark that in some models it is possible to construct a set-theoretic 
representation of contexts -- as some family of subsets of a set $\Omega.$ For example,
$\Omega$ can be the set of all possible parameters (e.g., physical, or mental, or economic)
of the model. However, in general we {\it do not assume the possibility to construct a set-theoretic 
representation of contexts.}

Another fundamental element of any contextual statistical model 
$M$ is a set of observables ${\cal O}:$
each observable $a\in {\cal O}$ can be measured
under each complex of conditions $C\in {\cal C}.$  
For an observable $a \in {\cal O},$ we denote the set
of its possible values (``spectrum'') by the symbol
$X_a.$

We do not assume that all these observables can  be measured
simultaneously. To simplify considerations, we shall consider only
discrete observables and, moreover, all concrete investigations will
be performed for {\it dichotomous observables.}

\medskip

{\bf Axiom 1:} {\it For  any observable
$a \in {\cal O}$  and its value $x \in X_a,$ there are defined contexts, say $C_x,$
corresponding to $x$-selections: if we perform a measurement of the observable $a$ under
the complex of physical conditions $C_x,$ then we obtain the value $a=x$ with
probability 1. We assume  that the set of contexts ${\cal C}$ contains 
$C_x$-selection contexts for all observables $a\in {\cal O}$ and $x\in X_a.$}

\medskip

For example, let $a$ be the observable corresponding to some question:  $a=+$ (the answer ``yes'')
and $a=-$ (the answer ``no''). Then the $C_{+}$-selection context is the selection of 
those participants of the experiment who answering ``yes'' 
to this question; in the same way we define the  
$C_{-}$-selection context. By Axiom 1 these contexts are well defined. 
We point out that in principle a participant of this experiment might not want to reply at all 
to this question.
By Axiom 1 such a possibility is excluded. By the same 
axiom both $C_{+}$ and $C_{-}$-contexts belong to the system of contexts under consideration.

\medskip

{\bf Axiom 2:} {\it There are defined contextual (conditional) 
probabilities ${\bf P}(a=x\vert C)$ for any
context $C \in {\cal C}$ and any observable $a \in {\it O}.$}

\medskip

Thus, for any context $C \in {\cal C}$ and any observable $a \in {\it O},$ 
there is defined the probability to observe the fixed value $a=x$ under the complex 
of conditions $C.$

Especially important role will be played by probabilities:
$$
p^{a\vert b}(x\vert y)\equiv {\bf P}(a=x\vert C_y), a, b \in {\cal O}, x \in X_a, y \in X_b,
$$
where $C_y$ is the $[b=y]$-selection context. By axiom 2 for any context $C\in {\cal C},$ 
there is defined the set of probabilities:
$$
 \{ {\bf P}(a=x\vert C): a \in {\cal O}\}.
$$
We complete this probabilistic data for the context $C$  by contextual probabilities with respect 
to the contexts $C_y$ corresponding to  the selections $[b=y]$ for all 
observables $b \in {\cal O}.$ The corresponding collection of data $D({\cal O}, C)$ 
consists of contextual probabilities:
$$
{\bf P}(a=x\vert C),{\bf P}(b=y\vert C),
{\bf P}(a=x\vert C_y), {\bf P}(b=y\vert C_x),...,
$$
where $a,b,... \in {\cal O}.$ Finally, we denote  the family of
probabilistic data $D({\cal O}, C)$ for all contexts  $C\in {\cal
C}$ by the symbol ${\cal D}({\cal O}, {\cal C}) 
(\equiv \cup_{C\in {\cal C}} D({\cal O}, C)).$

\medskip

{\bf Definition 1.} (V\"axj\"o Model) {\it An observational contextual  statistical model of reality is a triple
\begin{equation}
\label{VM}M =({\cal C}, {\cal O}, {\cal D}({\cal O}, {\cal C}))
\end{equation}
where ${\cal C}$ is a set of contexts and ${\cal O}$ is a  set of observables
which satisfy to axioms 1,2, and ${\cal D}({\cal O}, {\cal C})$ is probabilistic data
about contexts ${\cal C}$ obtained with the aid of observables belonging ${\cal O}.$}

\medskip

We call observables belonging the set ${\cal O}\equiv {\cal O}(M)$ {\it reference of observables.}
Inside of a model $M$  observables  belonging  to the set ${\cal O}$ give the only possible references
about a context $C\in {\cal C}.$

\section{Representation of Incomplete Information}

Probabilities ${\bf P}(b=y\vert C)$ are interpreted as {\it contextual (conditional) 
probabilities.}  We emphasize that we consider conditioning not with respect to {\it events}
as it is typically done in classical  probability \cite{Kolmogoroff}, but conditioning with 
respect to contexts -- complexes of (e.g., physical, biological, social,  mental, genetic, economic, or financial) 
conditions.  This is the crucial point. 

On the set of all events one can always introduce the structure of the {\it Boolean algebra}
(or more general $\sigma$-algebra). In particular, for any two events $A$ and $B$ 
their set-theoretic intersection $A \cap B$ is well defined and it determines a new event: 
the simultaneous occurrence of the events $A$ and $B.$ 

In contract to such an event-conditioning picture,
if one have two contexts, e.g., complexes of physical conditions $C_1$ and $C_2$ and if even 
it is possible to create the set-theoretic representation of contexts (as some 
collections of physical parameters), then, nevertheless, their set-theoretic intersection 
$C_1 \cap C_2$ (although it is well defined mathematically) need not correspond to 
any physically meaningful context. Physical contexts were taken just as examples. The same 
is valid for social, mental, economic, genetic and any other type of contexts.

Therefore even if for some model $M$ we can describe contexts in the set-theoretic 
framework, there are no reasons to assume that the collection of all contexts ${\cal C}$ 
should form a $\sigma$-algebra (Boolean algebra). This is the main difference from the classical 
(noncontextual) probability theory \cite{Kolmogoroff}.

One can consider the same problem from another perspective. Suppose that we have some 
set of parameters $\Omega$ (e.g., physical, or social, or mental). We also assume that contexts
are represented by some subsets of $\Omega.$ We consider two levels of description. At the
 first level a lot of information is available. There is a large set of contexts, we can even assume 
that they form a $\sigma$-algebra of subsets ${\cal F}.$ We call them the first level contexts.
There is a large number of observables at the first level, say the set of all possible 
random variables $\xi: \Omega \to {\bf R}$ (here ${\bf R}$ is the real line).
By introducing on ${\cal F}$ a probability measure ${\bf P}$ 
we obtain the classical Kolmogorov probability model $(\Omega, {\cal F}, {\bf P}),$
see \cite{Kolmogoroff}. This is the end of the classical story about the probabilistic 
description of reality. Such a model is used e.g. in classical statistical physics.

We point our that any Kolmogorov probability model induces  a V\"axj\"o model in such a way:
a) contexts are given by all sets $C\in {\cal F}$ such that ${\bf P}(C)\not=0;$
b) the set of observables coincides with the set of all possible random variables;
c) contextual probabilities are defined as Kolmogorovian conditional probabilities, i.e., 
by the Bayes formula: ${\bf P}(a=x\vert C)={\bf P}(\omega \in C: a(\omega)=x)/{\bf P}(C).$
This is the V\"axj\"o model for the first level of description. 

Consider now the second level of description. Here we can obtain 
a {\it non-Kolmogorovian V\"axj\"o model.} At this level only a part of information 
about the first level Kolmogorovian 
model $(\Omega, {\cal F}, {\bf P})$ can be obtained through a special family of 
observables ${\cal O}$ which correspond to a special subset of the set of all random variables 
of the Kolmogorov model $(\Omega, {\cal F}, {\bf P})$ at the first level of description.
Roughly speaking not all contexts of the first level, 
${\cal F}$ can be ``visible'' at the second level. There is no sufficiently many 
observables ``to see'' all contexts of the first level -- elements of the 
Kolmogorov $\sigma$-algebra ${\cal F}.$ Thus we should cut off this $\sigma$-algebra 
${\cal F}$ and obtain
a smaller family, say ${\cal C},$ of visible contexts. Thus some V\"axj\"o models (those permitting 
a set-theoretic representation) can appear 
starting with the purely classical Kolmogorov probabilistic framework, as a consequence
of ignorance of  information. If not all information is available, so we cannot use the 
first level (classical) description, then we, nevertheless, can proceed with the second level 
contextual description.

We shall see that starting with some V\"axj\"o models we can obtain the 
quantum-like calculus of probabilities in the complex Hilbert space. Thus in the opposition 
to a rather common opinion, we can derive a {\it quantum-like description 
for ordinary macroscopic systems}
as the results of using of an incomplete representation. 
This opens great possibilities in application
of quantum-like models outside the micro-world. In particular, in cognitive science 
we need not consider composing of the brain from quantum particles to come 
to the quantum-like model.

\medskip

{\bf Example 1.} (Firefly in the box) Let us consider a box which is divided into 
four sub-boxes. These small boxes which are denoted by 
$\omega_1, \omega_2, \omega_3, \omega_4$ provides
the the first level of description. We consider a Kolmogorov probability space:
$\Omega=\{ \omega_1, \omega_2, \omega_3, \omega_4\},$ the algebra of all finite subsets 
${\cal F}$ of $\Omega$ and a probability measure determined by probabilities 
${\bf P}(\omega_j)=p_j,$ where $ 0< p_j< 1, p_1+...+p_4 = 1.$ We remark that in 
our interpretation it is more natural to consider elements of $\Omega$ as 
{\it elementary parameters,} and not as {\it elementary events} (as it was done by Kolmogorov). 

\psset{unit=1cm}

\begin{figure}
\centering
\begin{pspicture}(0,0)(2,2)
\psframe(0,0)(2,2)
\psline(0,1)(2,1)
\psline(1,0)(1,2)

\rput(0.5, 1.3){\parbox{1.75cm} {\raggedright\small
\begin{center}$\omega_1$\end{center}}}
\rput(1.5, 1.5){\parbox{1.75cm} {\raggedright\small \begin{center}$\omega_3$
\end{center}}}
\rput(0.5, 0.5){\parbox{1.75cm} {\raggedright\small \begin{center}$\omega_2$
\end{center}}}
\rput(1.5, 0.5){\parbox{1.75cm} {\raggedright\small \begin{center}$\omega_4$
\end{center}}}

\end{pspicture}
\caption{The first level (complete) description.}
\end{figure}

We now consider two different disjoint partitions of the set $\Omega:$ 
$$
A_1=\{ \omega_1, \omega_2\}, A_2=\{ \omega_3, \omega_4\},
$$
$$
B_1=\{\omega_1, \omega_4\}, B_1=\{\omega_2, \omega_3\}.
$$
We can obtain such partitions by dividing the box: 
a) into two equal parts by the vertical line:
the left-hand part gives $A_1$ and the right-hand part $A_2;$
b) into two equal parts by the horizontal line:
the top part gives $B_1$ and the bottom part $B_2.$

We introduce two  random variables corresponding to these partitions:
$\xi_a(\omega) =x_i,$ if $\omega\in A_i$ and $\xi_b(\omega)=y_i\in$ if $\omega\in B_i.$ 
Suppose now that we are able to measure only these two variables, denote the corresponding
observables by the symbols $a$ and $b.$ We project the Kolmogorov model under consideration to 
a non-Kolmogorovian V\"axj\"o model by using the observables $a$ and $b$ -- the second 
level of description. At this level the set of observables ${\cal O}=\{ a, b\}$ and 
the natural set of contexts ${\cal C}: \Omega, A_1, A_2,
 B_1, B_2, C_1=\{\omega_1,\omega_3\},
C_1=\{\omega_2,\omega_4\}$ and all unions of these sets. Here ``natural'' has the meaning permitting 
a quantum-like representation (see further considerations). Roughly speaking contexts of the second 
level of description should be large enough to ``be visible'' with the aid 
of observables $a$ and $b.$ 

\psset{unit=1cm}

\begin{figure}
\centering
\begin{pspicture}(0,0)(2,2)
\psframe(0,0)(2,2)
\psline(1,0)(1,2)
\rput(0.5, 1.3){\parbox{1.75cm} {\raggedright\small
\begin{center}$A_1$\end{center}}}
\rput(1.5, 1.5){\parbox{1.75cm} {\raggedright\small \begin{center}$A_2$
\end{center}}}
\psdot(1, 0)
\end{pspicture}
\caption{The second level description: the a-observable.}
\end{figure}
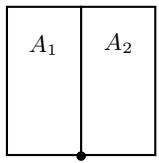

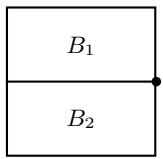
\begin{figure}
\centering
\begin{pspicture}(0,0)(2,2)
\psframe(0,0)(2,2)

\psline(0,1)(2,1)
\psdot(2,1)
\rput(1, 1.3){\parbox{1.75cm} {\raggedright\small
\begin{center}$B_1$\end{center}}}
\rput(1, 0.5){\parbox{1.75cm} {\raggedright\small \begin{center}$B_2$
\end{center}}}
\end{pspicture}
\caption{The second level description: the b-observable}
\end{figure}

Intersections of these sets need not belong 
to the system of contexts (nor complements of these sets). Thus the Boolean structure of 
the original first level description disappeared, but, nevertheless, it is present in the latent form. 
Point-sets $\{ \omega_j \}$ are not  ``visible'' at this level of description. 
For example, the random variable $$
\eta(\omega_j)= \gamma_j, j=1,..., 4, \; \gamma_i \not= \gamma_j, i\not=j,
$$ 
is not an observable at the second level. 

Such a model was discussed from positions of quantum logic, see, e.g., \cite{Svozil}.
There can be provided a nice interpretation of these two levels of description. Let us consider
a firefly in the box. It can fly everywhere in this box. Its locations are described by 
the uniform probability distribution ${\bf P}$ (on the $\sigma$-algebra of Borel 
subsets of the box).  This is the first level of description. Such a description 
can be realized if the box were done from glass or if at every point of the 
box there were  a light detector. All Kolmogorov random variables can be considered as observables. 

Now we consider the situation when there are only two possibilities 
to observe the firefly in the box: 

\medskip

1) to open a small window at a point $a$ which is located in 
such a way (the bold dot in the middle of the bottom side of the box) that it is possible to determine  only either the firefly is in the section $A_1$ 
or in the section $A_2$ of the box; 

2) to open a small window at a point $b$ which is located in 
such a way (the bold dot in the middle of the right-hand side of the box) that it is possible to determine  only either the firefly is in the section $B_1$ 
or in the section $B_2$ of the box. 

\medskip

In the first case I can determine in which part, $A_1$ or $A_2,$
the firefly is located. In the second case I also can only determine in which part, 
$B_1$ or $B_2,$ the firefly is located. But I am not able to look into both windows
simultaneously. In such a situation the observables $a$ and $b$ 
are the only source of information about the firefly (reference observables). 
The Kolmogorov description is meaningless (although it is incorporated in the model 
in the latent form). Can one apply a quantum-like  description, namely, represent 
contexts  by complex probability amplitudes? The answer is to be positive.
The set of contexts that permit 
the quantum-like representation consists of all subsets $C$ such that ${\bf P}(A_i\vert C)>0$ and 
${\bf P}(B_i\vert C)>0, i=1,2$  (i.e., for sufficiently large contexts). We have 
seen that the Boolean 
structure disappeared as a consequence of ignorance of information. 

\medskip

Finally, we emphasize again that the V\"axj\"o model is essentially more general.  
The set-theoretic representation need not exist at all.

\section{Quantum Projection of Boolean Logic}

Typically the absence of the Boolean structure on the set of quantum propositions is 
considered as the violation of
laws of classical logic, e.g., in quantum mechanics \cite{Birkhoff/vonNeumann}. 
In our approach classical logic is 
not violated, it is 
present in the latent form. However, we are not able to use it, because we do not have complete
information. Thus quantum-like logic is a kind of projection of classical logic. The impossibility 
of
operation with complete information about a system is not always a disadvantages. Processing 
of incomplete
set of information has the evident advantage comparing with ``classical Boolean'' complete information 
processing -- the great saving of computing resources and increasing of the speed of computation.
However, the Boolean structure cannot be violated in an arbitrary way, because in such a case 
we shall get a chaotic computational process. There should be developed some calculus of consistent 
ignorance by information. Quantum formalism provides one of such calculi.

Of course, there are no reasons to assume 
that processing of information through ignoring of its essential part should be rigidly coupled 
to a special class of physical systems, so called quantum systems. Therefore we prefer to speak about
{\it quantum-like processing of information} that may be performed by various kinds of 
physical and biological systems. In our approach quantum computer has advantages
not because it is based on a special class of physical systems (e.g., electrons or ions), but because
there is realized the consistent processing of incomplete information. We prefer to use the 
terminology $QL$-computer by reserving the ``quantum computer'' for a special class of QL-computers
which are based on quantum physical systems. 

One may speculate that some biological systems could develop in the process of evolution the 
possibility to operate in a consistent way with incomplete information. Such a QL-processing of 
information implies evident advantages. Hence, it might play an important role in the process
of the natural selection. It might be that consciousness is a form of the QL-presentation 
of information. In such a way we really came back to Whitehead's analogy between quantum and
conscious systems \cite{Whitehead}.

\section{Contextual Interpretation of ``Incompatible'' Observables}

Nowadays the notion of incompatible (complementary) observables is rigidly coupled to 
{\it noncommutativity.} In the 
conventional quantum formalism observables are incompatible  iff they are represented by 
noncommuting
self-adjoint operators $\hat{a}$ and $\hat{b}:  [\hat{a}, \hat{b}]\not=0.$ As we see, the V\"axj\"o 
model is not from the very beginning coupled to a representation of information in a Hilbert space.
Our aim is to generate an analogue (may be not direct) of the notion of incompatible (complementary)
observables starting not from the mathematical
formalism of quantum mechanics, but  on the basis of the V\"axj\"o model, i.e., directly
from statistical data.

Why do I dislike the conventional identification of {\it incompatibility 
with noncommutativity?} 
The main reason is that typically the mathematical formalism of quantum mechanics is identified
with it as a physical theory. Therefore the quantum incompatibility  represented
through noncommutativity 
is rigidly coupled
to the micro-world. (The only possibility to transfer quantum behavior to the macro-world is to consider
physical states of the Bose-Einstein condensate type.)  We shall see that some V\"axj\"o models
can be represented as the conventional quantum model in the complex Hilbert space. However, the 
V\"axj\"o model is essentially more general than the quantum model. In particular, some V\"axj\"o 
models can be represented not in the complex, but in hyperbolic Hilbert space (the Hilbert module
over the two dimensional Clifford algebra with the generator $j: j^2=+1).$

Another point is that the terminology -- incompatibility  -- is misleading in our approach. 
The quantum mechanical meaning of compatibility is the possibility to measure two observables,
$a$ and $b$ {\it simultaneously.} In such a case they are represented by commuting operators.
Consequently incompatibility implies the impossibility of  simultaneous measurement 
of $a$ and $b.$ In the V\"axj\"o model there is no such a thing as fundamental impossibility
of simultaneous measurement. We present the viewpoint that quantum  incompatibility is just a 
consequence of information {\it supplementarity} of observables $a$ and $b.$ The information 
which is obtained via a measurement of, e.g., $b$ can be non trivially updated by additional 
information which is contained in the result of a measurement of $a.$ Roughly speaking if one knows 
a value of $b,$ say $b=y,$ this does not imply knowing the fixed value of $a$ 
and vice versa, see \cite{Khrennikov/Supp} for details.

We remark that it might be better to use the notion ``complementary,''
instead of ``supplementary.'' However, the first one was already reserved by Nils Bohr for 
the notion which very close to ``incompatibility.'' In any event Bohr's complementarity implies 
{\it mutual exclusivity} that was not the point of our considerations. 

Supplementary processes 
take place
not only in physical micro-systems. For example,  in the brain 
there are present supplementary mental processes. Therefore the brain is a (macroscopic)
QL-system. Similar supplementary processes take place in economy and in particular 
at financial market. There one could also use quantum-like descriptions \cite{Choustova}. 
But the essence
of the quantum-like descriptions is not the representation of information in the 
complex Hilbert space, but incomplete (projection-type) representations of information. 
It seems that the V\"axj\"o model provides a rather general description of such representations.

We introduce a notion of supplementary which will produce in some cases
the quantum-like representation of observables by noncommuting operators, but which is not 
identical to incompatibility (in the sense of impossibility of simultaneous observations) nor 
complementarity (in the sense of mutual exclusivity).

\medskip

{\bf Definition 2.} {\it Let  $a,b \in {\cal O}.$ The observable $a$  is said to be
supplementary  to the observable $b$ if
\begin{equation}
\label{VM5}
p^{a\vert b}(x\vert y)\not= 0, 
\end{equation}
for all $x \in X_a, y\in X_b.$}

\medskip

Let $a=x_1, x_2$ and $b=y_1, y_2$ be two dichotomous observables.
In this case  (\ref{VM5}) is equivalent  to the condition:
\begin{equation}
\label{VM5Z}
p^{a\vert b}(x\vert y)\not= 1,
\end{equation}
for all $x \in X_a, y\in X_b.$
Thus by knowing the result $b=y$ of the $b$-observation we are not able to make the definite
prediction about the result of the $a$-observation. 

Suppose now that  (\ref{VM5Z}) is violated (i.e., $a$ is not supplementary to $b),$ for example:
\begin{equation}
\label{VM5Z1}
p^{a\vert b}(x_1\vert y_1) = 1,
\end{equation}
and, hence, 
$p^{a\vert b}(x_2\vert y_1) = 0.$
Here the result $b=y_1$ determines the result $a=x_1.$

In future we shall consider a special class of V\"axj\"o models in that the matrix of transition 
probabilities ${\bf P}^{a\vert b} =(p^{a\vert b}(x_i\vert y_j))_{i,j=1}^2$ is {\it double
stochastic:} 
$
p^{a\vert b}(x_1\vert y_1) +p^{a\vert b}(x_1\vert y_2)=1;
p^{a\vert b}(x_2\vert y_1) +p^{a\vert b}(x_2\vert y_2)=1.
$
In such a case the condition (\ref{VM5Z1}) implies that 
\begin{equation}
\label{VM5Z3}
p^{a\vert b}(x_2\vert y_2) = 1,
\end{equation}
and, hence, 
$p^{a\vert b}(x_1\vert y_2) = 0.$
Thus also the result $b=y_2$ determines the result $a=x_2.$

We point out that for models with double stochastic matrix 
$${\bf P}^{a\vert b} =(p^{a\vert b}(x_i\vert y_j))_{i,j=1}^2$$ the relation of 
supplementary is symmetric! In general it is not the case. It can happen that $a$ is 
supplementary to $b:$ each $a$-measurement gives us additional information updating
information obtained in a preceding measurement of $b$ (for any result $b=y).$ But 
$b$ can be non-supplementary to $a.$ 

Let us now come back to Example 1. The observables $a$ and $b$ are supplementary in our meaning.
Consider now the classical Kolmogorov model and suppose that we are able to measure not only 
the random variables $\xi_a$ and $\xi_b$ -- observables $a$ and $b,$ but also the random variable
$\eta.$ We denote the corresponding observable by $d.$ 
The pairs of observables $(d,a)$ and $(d,b)$ are non-supplementary:
\[p^{a\vert d}(x_1\vert \gamma_i) = 0, \; i= 3,4;\; 
p^{a\vert d}(x_2\vert \gamma_i) = 0, \; i= 1,2,\]
and, hence,
\[p^{a\vert d}(x_1\vert \gamma_i) = 1, \; i= 1,2;\;
p^{a\vert d}(x_2\vert \gamma_i) = 1, \; i= 3,4.\]
Thus if one knows , e.g., that $d=\gamma_1$ then 
it is definitely that $a=x_1$ and so on.

\section{A Statistical Test to  Find Quantum-like Structure}

 We consider examples of  cognitive contexts: 

1). $C$ can be some selection procedure that is used to select a special group $S_C$ of people 
or animals.
 Such a context is represented by this group $S_C$ (so this is an ensemble of cognitive systems). 
 For example, we select a group $S_{\rm{prof.math.}}$
 of professors of mathematics
 (and then ask questions $a$ or (and) $b$ or give corresponding tasks).  We can select a group of
 people of some age. We can select a group of people having a ``special mental state'':
 for example, people in love  or hungry people (and then ask questions or give tasks).

2). $C$ can be a learning procedure that is used to create some
special group of people or animals.  For example, rats can be
trained to react to  special stimulus.
 
We can also consider {\it social contexts.} For example, social classes: proletariat-context, bourgeois-context;
 or war-context, revolution-context, context of economic depression, poverty-context,
 and so on. Thus our model can be used
  in social and political sciences (and even in history). We can try to find quantum-like statistical
  data in these sciences.

We describe a mental interference experiment.

Let $a=x_1, x_2$ and $b=y_1, y_2$ be two dichotomous mental observables:
$x_1$=yes, $x_2$=no, $y_1$=yes, $y_2$=no.
We set $X\equiv X_a= \{x_1, x_2\}, Y\equiv X_b= \{y_1,y_2\}$ 
(``spectra'' of observables $a$ and $b).$
Observables can be two  different questions or two different types of cognitive tasks.
We use these two fixed reference observables for probabilistic 
representation of cognitive contextual reality given by $C.$

We perform observations of $a$ under the complex of cognitive conditions $C:$   
 $$
p^a(x)= \frac{\mbox{the number of results}\; a=x}{\mbox{the total number of observations}}.
 $$
 So $p^a(x)$ is the probability to get the result $x$ for observation of the $a$
 under the complex of cognitive conditions $C.$
 In the same way we find probabilities $p^b(y)$ for the $b$-observation  under the same cognitive context $C.$

As was supposed in axiom 1, cognitive contexts $C_y$
can be created corresponding to selections with respect
to fixed values of the $b$-observable.  The context $C_y$ (for fixed $y \in Y)$ can be characterized in the following way. By measuring the $b$-observable under the cognitive context $C_y$
we shall obtain the answer $b= y$ with probability one.
We perform now the $a$-measurements under cognitive contexts $C_y$ for $y= y_1, y_2,$
and find the probabilities:
\[p^{a\vert b}(x\vert y)=\frac{\mbox{number of the result} \; a=x \;\mbox{for context} \;  C_y}
{\mbox{number of all observations} \;\mbox{for context} \;  C_y}\]
where $x \in X, y \in Y.$ For example, by using the ensemble approach to probability 
we have that the probability $p^{a\vert b}(x_1\vert y_2)$ is obtained
as the frequency of the answer $a=x_1=yes$ in the ensemble of cognitive system that have already
answered $b=y_2=no.$  Thus we first select a sub-ensemble of cognitive systems who 
replies $no$ to the $b$-question, $C_{b=no}.$ Then we ask systems belonging to 
$C_{b=no}$ the $a$-question.

 It is assumed (and this is a very natural assumption) that a cognitive system is
 ``responsible for her (his) answers.'' Suppose that a system $\tau$ has answered $b=y_2=no.$
 If we ask $\tau$ again the same question $b$ we shall
 get the same answer $b=y_2=no.$  This is nothing else than {\it the mental form of the
 von Neumann projection postulate:} the second
measurement of the same observable, performed immediately after the
first one, will yield the same value of the observable).

Classical probability theory tells us that all these probabilities have to be connected by the so
 called {\it formula of total probability:}
 \[p^a(x)= p^b(y_1)p^{a\vert b}(x\vert y_1) + p^b(y_2) p^{a\vert b}(x\vert y_2), \;\; x \in X .\]
 However, if the theory is quantum-like, then we should obtain  \cite{Khrennikov/book:2004} the
 formula of total probability with an interference term:
 \begin{equation}
 \label{LLL}
 p^a(x)= p^b(y_1)p^{a\vert b}(x\vert y_1) + p^b(y_2) p^{a\vert b}(x\vert y_2)
 \end{equation}
 \[+ 2  \lambda(a=x\vert b, C) \sqrt{p^b(y_1)p^{a\vert b}(x\vert y_1)p^b(y_2) p^{a\vert b}(x\vert y_2) },\]
 where the coefficient of supplementarity (the coefficient of interference) is given by 
\begin{equation}
 \label{LLLT}
\lambda(a=x\vert b, C)= \frac{ p^a(x)- p^b(y_1)p^{a\vert b}(x\vert y_1) - p^b(y_2) p^{a\vert b}(x\vert y_2) }
  {2\sqrt{p^b(y_1)p^{a\vert b}(x\vert y_1)p^b(y_2) p^{a\vert b}(x\vert y_2)}}
\end{equation}
This formula holds true for {\it supplementary observables.} To prove its validity, it is sufficient to put
the expression for $\lambda(a=x\vert b, C),$ see (\ref{LLLT}), into (\ref{LLL}).
In the  quantum-like statistical test for a cognitive context $C$ we
 calculate 
$$\bar{\lambda}(a=x\vert b, C)=
 \frac{ \bar{p}^a(x)- \bar{p}^b(y_1)\bar{p}^{a\vert b}(x\vert y_1) - \bar{p}^b(y_2) 
\bar{p}^{a\vert b}(x\vert y_2) }
{2\sqrt{\bar{p}^b(y_1)\bar{p}^{a\vert b}(x\vert y_1)\bar{p}^b(y_2) \bar{p}^{a\vert b}(x\vert y_2)}},
$$
where the symbol $\bar{p}$ is used for empirical probabilities -- frequencies.
An empirical situation with $ \lambda(a=x\vert b, C) \not =0$ would yield evidence for quantum-like
behaviour of cognitive systems. In this case, starting with (experimentally calculated) coefficient
of interference $\lambda(a=x\vert b, C)$
we can proceed either to the conventional Hilbert space formalism (if this coefficient is bounded by 1)
or to so called hyperbolic Hilbert space formalism (if this coefficient is larger than  1).
In the first case the coefficient of interference can be represented in the trigonometric form
 $
 \lambda(a=x\vert b, C)= \cos \theta(x),
 $
 Here $\theta(x)\equiv \theta(a=x\vert b, C)$ is the phase of the $a$-interference between
 cognitive contexts $C$ and $C_y, y \in Y.$ In this case we have the conventional formula of total probability with the interference term:
  \begin{equation}
 \label{LLLQ}
 p^a(x)= p^b(y_1)p^{a\vert b}(x\vert y_1) + p^b(y_2) p^{a\vert b}(x\vert y_2)
 \end{equation}
 \[+ 2  \cos \theta(x) \sqrt{p^b(y_1)p^{a\vert b}(x\vert y_1)p^b(y_2) p^{a\vert b}(x\vert y_2) }.\]
 In principle, it could be derived in the conventional Hilbert space formalism. 
But we chosen the inverse way.
Starting with (\ref{LLLQ}) we could introduce a ``mental wave function''
 $\psi\equiv \psi_C$ (or pure quantum-like mental state) belonging to 
this Hilbert space.
 We recall that in our approach
 a mental wave function $\psi$  is just a representation of a cognitive context $C$
by a complex probability amplitude. The latter provides a Hilbert representation of 
statistical data about context which can be obtained with the help of two fixed observables 
(reference observables).

\section{The Wave Function Representation of Contexts}

In this section we shall present an algorithm for representation of a context (in fact, 
probabilistic data on this context) by a complex probability amplitude. 
This QL representation algorithm (QLRA) was created by the author \cite{Khrennikov/book:2004}. 
It can be interpreted 
as a consistent projection of classical probabilistic description (the complete one) 
onto QL probabilistic description (the incomplete one). 
 
Let $C$ be a context. We consider only  contexts
with trigonometric interference for {\it supplementary 
observables} $a$ and $b$ -- the reference observables for coming QL  
representation of contexts. The collection of all trigonometric contexts, i.e., contexts having the coefficients
of supplementarity (with respect to two fixed observables $a$ and $b$) bounded by one, 
is denoted by the symbol  ${\it C}^{\rm{tr}}\equiv{\it C}^{\rm{tr}}_{a\vert b}.$
We again point out to the dependence of the notion of a trigonometric context on the 
choice of reference observables.

We now point directly to dependence of probabilities on contexts by using 
the context lower index $C.$ The interference formula 
of total probability (\ref{LLL}) can be written
in the following form: 
\begin{equation}
\label{Two}
p_{C}^a(x)= \sum_{y \in X_b} p_{C}^b(y) p^{a\vert b}(x\vert y) +
2\cos \theta_{C}(x)\sqrt{\Pi_{y \in  Y}p_{C}^b(y) p^{a\vert b}(x\vert y)}
\end{equation}

By using the elementary formula: 
$$D=A+B+2\sqrt{AB}\cos
\theta=\vert \sqrt{A}+e^{i \theta}\sqrt{B}|^2,$$ for $A, B > 0,
\theta\in [0,2 \pi],$ we can represent the probability $p_C^a(x)$ as
the square of the complex amplitude (Born's rule):
\begin{equation}
\label{Born} p_C^a(x)=\vert\varphi_C(x)\vert^2 ,
\end{equation}
where a complex probability amplitude is defined  by

\begin{equation}
\label{EX1} 
\varphi(x)\equiv \varphi_C(x) =\sqrt{p_C^b(y_1)p^{a\vert b}(x \vert y_1)}
+ e^{i \theta_C(x)} \sqrt{p_C^b(y_2)p^{a \vert b}(x \vert y_2)} \;.
\end{equation}

We denote the space of functions: $\varphi: X_a \to {\bf C},$
where ${\bf C}$ is the field of complex numbers, by the
symbol $\Phi =\Phi(X_a, {\bf C}).$ Since $X_a= \{x_1, x_2 \},$ the
$\Phi$ is the two dimensional complex linear space. By using the
representation (\ref{EX1}) we construct the map 
$$
J^{a \vert b}: {\it C}^{\rm{tr}} \to \Phi(X, {\bf C})
$$ 
which maps contexts (complexes of, e.g.,
physical or social conditions) into complex amplitudes.
This map realizes QLRA.

 The representation
({\ref{Born}}) of probability is nothing other than the famous {\bf
Born rule.} The complex amplitude $\varphi_C(x)$ can be called a
{\bf wave function} of the complex of physical conditions (context)
$C$  or a  (pure) {\it state.}  We set $e_x^a(\cdot)=\delta(x-
\cdot).$ The Born's rule for complex amplitudes (\ref{Born}) can be
rewritten in the following form: 
\begin{equation}
\label{BH}
p_C^a(x)=\vert(\varphi_C, e_x^a)\vert^2,
\end{equation}
where the scalar product
in the space $\Phi(X, C)$ is defined by the standard formula:
\begin{equation}
\label{RRRR}
(\varphi, \psi) = \sum_{x\in X_a} \varphi(x)\bar \psi(x).
\end{equation}
 The system
of functions $\{e_x^a\}_{x\in X_a}$ is an orthonormal basis in the
Hilbert space $$
{\bf H}=(\Phi, (\cdot, \cdot)).
$$ 
Let $X_a$ be a subset of the real line. By using
the Hilbert space representation  of the Born's rule  we
obtain  the Hilbert space representation of the expectation of the
reference observable $a$: 
$$
E (a\vert C)= \sum_{x\in
X_a}x\vert\varphi_C(x)\vert^2= \sum_{x\in X_a}x (\varphi_C, e_x^a)
\overline{(\varphi_C, e_x^a)}= (\hat a \varphi_C, \varphi_C),
$$ 
where the  (self-adjoint) operator $\hat a: {\bf H} \to {\bf H}$ is determined by its
eigenvectors: $\hat a e_x^a=x e^a_x, x\in X_a.$ This is the
multiplication operator in the space of complex functions
$\Phi(X,{\bf C}):$ 
$$
\hat{a} \varphi(x) = x \varphi(x). 
$$ 
It is natural to represent this reference observable (in the Hilbert space
model)  by the operator $\hat a.$

We would like to have Born's rule not only for the $a$-observable, but also for the $b$-observable.
$$
p_C^b(y)=\vert(\varphi, e_y^b)\vert^2 \;, y \in  X_b.
$$
Thus both reference observables would be represented by self-adjoint operators determined by bases
$\{e_x^a\}, \{e_y^b\},$ respectively.

How can we define the basis $\{e_y^b\}$ corresponding to the
$b$-observable? Such a basis can be found starting with interference
of probabilities. We set $u_j^b=\sqrt{p_C^b(y_j)},
p_{ij}=p^{a\vert b}(x_j \vert y_i), u_{ij}=\sqrt{p_{ij}}, \theta_j=\theta_C(x_j).$ We
have:
\begin{equation}
\label{0} \varphi=u_1^b e_1^b + u_2^b e_2^b,
\end{equation}
where
\begin{equation}
\label{Bas} e_1^b= (u_{11}, \; \; u_{12}) ,\; \; e_2^b= (e^{i
\theta_1} u_{21}, \; \; e^{i \theta_2} u_{22})
\end{equation}
We consider the {\it matrix of transition probabilities} ${\bf
P}^{a\vert b}=(p_{ij}).$ It is always a  {\it stochastic matrix:}
$p_{i1}+p_{i2}=1, i=1,2).$ We remind  that a matrix is called  {\it
double stochastic} if it is stochastic and, moreover, $p_{1j} +
p_{2j}=1, j=1,2.$ The  system $\{e_i^b\}$   is an orthonormal basis iff the matrix ${\bf
P}^{a\vert b}$ is double stochastic and probabilistic phases satisfy the
constraint: $ \theta_2 - \theta_1= \pi \; \rm{mod} \; 2 \pi,$ see \cite{Khrennikov/book:2004} 
for details.

It will be always supposed that the matrix of transition probabilities 
${\bf P}^{a\vert b}$ is double
stochastic. In this case the $b$-observable is represented by the operator
$\hat{b}$ which is diagonal (with eigenvalues $y_i)$ in the basis
$\{e_i^b\}.$ The Kolmogorovian conditional average of the random
variable $b$ coincides with the quantum Hilbert space average:
$$
E(b\vert C)=\sum_{y \in X_b} y p_C^b(y) = (\hat{b} \phi_C, \phi_C), \; C \in {\it
C}^{\rm{tr}}.
$$

\section{Brain as a System Performing a Quantum-like Processing of Information}

The brain is a huge information system that contains millions of
elementary mental states . It could not ``recognize'' (or ``feel'') all those states  at
each instant of time $t.$ Our fundamental hypothesis is that the
brain is able to create the QL-representations of mind. At each
instant of time $t$ the brain creates the  QL-representation of its
mental context $C$ based on two supplementary mental
(self-)observables $a$ and $b.$ Here $a=(a_1,..., a_n)$ and
$b=(b_1,..., b_n)$ can be very long vectors of compatible (non-supplementary)
dichotomous observables. The reference observables $a$ and $b$ can be
chosen (by the brain) in different ways at different instances of
time. Such a change of the reference observables is known in
cognitive sciences as a {\it change of representation.}

A mental context $C$ in the $a\vert b-$ representation is  described
by the mental wave function $\psi_C.$ We can speculate that the
brain has the ability to feel this mental field as a distribution on
the space $X.$ This distribution is given by the norm-squared of the
mental wave function: $\vert \psi_C(x) \vert^2.$

In such a model it might be supposed that the state of our
consciousness is represented by the mental wave function $\psi_C.$
The crucial point is that in this model consciousness is created
through neglecting an essential volume of information contained in
subconsciousness. Of course, this is not just a random loss of
information. Information is selected through QLRA, see (\ref{EX1}): a mental
context $C$ is projected onto the complex probability amplitude $\psi_C.$

The (classical) mental state of sub-consciousness evolves with time
$C\to C(t).$ This  dynamics induces dynamics of the mental wave
function $\psi(t)=\psi_{C(t)}$ in the complex Hilbert space.

Further development of our approach (which we are not able to
present here) induces the following model of brain's functioning \cite{Khrennikov/QLBrain}:

{\it The brain is able to create the QL-representation of mental
contexts, $C\to \psi_C$ (by using the algorithm based on the formula
of total probability with interference).}

\section{Brain as Quantum-like Computer}

The ability of the brain to create 
the QL-representation of mental contexts
 induces functioning of the brain as a quantum-like computer.

\medskip

{\it The brain performs computation-thinking by using algorithms of quantum computing
in the complex Hilbert space of mental QL-states.}

\medskip

We emphasize that in our approach the brain is not  quantum computer, but a QL-computer. On one hand, a QL-computer
works totally in accordance with the mathematical theory of quantum computations (so by using quantum algorithms).
 On the other hand, it is not based on superposition of individual mental states. The complex amplitude
  $\psi_C$ representing a mental context $C$ is a special probabilistic representation of information states
  of the huge neuronal ensemble.
In particular, the brain is a {\it macroscopic} QL-computer. Thus the QL-parallelism (in the opposite to
conventional quantum parallelism) has a natural realistic base. This is real parallelism in the working
of millions of neurons. The crucial point is the way in which this classical parallelism is projected
onto dynamics of QL-states. The QL-brain is able to solve  $NP$-problems. But there is nothing mysterious
in this ability: an exponentially increasing number of operations is
performed through involving of an exponentially increasing number of neurons.

We point out that by coupling QL-parallelism to working of neurons we started to present a particular ontic model
for QL-computations. We shall discuss it in more detail. Observables $a$ and $b$ are self-observations of the brain. 
They can be represented as functions of the internal state of brain $\omega.$ Here $\omega$ is a parameter of huge dimension
describing states of all neurons in the brain: $\omega= (\omega_1, \omega_2,..., \omega_N):$
$$
a=a(\omega), b=b(\omega).
$$
The brain is not interested in concrete values of the reference observables at fixed instances of time. The brain
finds the contextual probability distributions $p_C^a(x)$ and $p_C^b(y)$ and creates the mental QL-state $\psi_C(x),$
see QLRA -- (\ref{EX1}).
Then it works with the mental wave function $\psi_C(x)$ by using algorithms of quantum computing. 

\section{Two Time Scales as the Basis of the QL-representation of Information}

The crucial problem is to find  a mechanism
for producing contextual probabilities. 
We think that they are frequency probabilities that are created in
 the brain in the following way.There are two scales of time: 
a) internal scale, $\tau$-time; 
b) QL-scale, $t$-time.
The internal scale is {\it finer than the QL-scale.}
Each instant of QL-time $t$ corresponds to an 
interval $\Delta$ of internal time $\tau.$ We might 
identify the QL-time with mental (psychological) time and 
the internal time with physical time. We shall also use the terminology:
pre-cognitive time-scale - $\tau$ and cognitive time-scale - $t.$ 

During the interval $\Delta$ of internal time the brain 
collects statistical data for self-observations of $a$ and $b.$
The internal state $\omega$ of the brain evolves as 
$$\omega= \omega(\tau, \omega_0).$$ 
This is a classical dynamics (which can be described by a 
stochastic differential equation).

At each instance of internal time $\tau$ there are 
performed nondisturbative self-measurements of $a$ and $b.$ 
These are realistic measurements: the brain gets values 
$a(\omega(\tau, \omega_0)),$ $b(\omega(\tau, \omega_0)).$ 
By finding frequencies of realization of fixed values for 
$a(\omega(\tau, \omega_0))$ and  $b(\omega(\tau, \omega_0))$ 
during the interval $\Delta$ of internal time,
the brain obtains the frequency probabilities
$p_C^a(x)$ and $p_C^b(y).$ These probabilities are related to 
the instant of QL-time time $t$ corresponding to the interval of internal time $\Delta:$
$p_C^a(t, x)$ and $p_C^b(t,y).$  We remark that in these probabilities the brain encodes huge 
amount of information -- millions of mental ``micro-events'' which happen during the 
interval $\Delta.$ But the brain is not interested in all those individual events. (It would be too
disturbing and too irrational to take into account all those fluctuations of mind.) It takes into 
account only the integral result of such a {\it pre-cognitive activity} (which was performed at the
pre-cognitive time scale).

For example, the mental observables $a$ and $b$ can 
be measurements over different  domains of brain. It is supposed that the brain
 can ``feel'' probabilities
(frequencies) $p_C^a(x)$ and $p_C^b(y),$ but not able to ``feel'' 
the simultaneous probability distribution
$p_C(x,y) = P(a=x, b=y\vert C).$ 

This is not the problem of mathematical existence of such a distribution.
This is the problem of integration of statistics of observations from 
different domains of the brain.
By using the QL-representation based only on probabilities 
$p_C^a(x)$ and $p_C^b(y)$  the brain could be able to
escape integration of information about {\it individual self-observations} of variables $a$ and $b$
related to spatially separated domains of brain. The brain  need not couple 
these domains at each instant of internal (pre-cognitive time)  time $\tau.$ 
It couples them only once in the interval $\Delta$ through the contextual probabilities
$p_C^a(x)$ and $p_C^b(y).$ This induces the huge saving of time and increasing of speed of 
processing of mental information. 

One of fundamental consequences for cognitive science is that our mental images have the 
probabilistic structure. They are products of transition from an extremely fine 
pre-cognitive time scale to a rather rough cognitive time scale.

Finally, we remark that a similar time scaling approach was developed in \cite{Khrennikov/T} 
for ordinary quantum 
mechanics. In \cite{Khrennikov/T} quantum expectations appear as results of averaging with 
respect to a prequantum 
time scale. There was presented an extended discussion of possible 
choices of quantum and prequantum time scales. 

 We can discuss the same problem in the cognitive framework.
We may try to estimate the time scale parameter
$\Delta$ of  the neural QL-coding. There are strong experimental evidences,
see, e.g.,  \cite{Mori}, that a moment in psychological time correlates 
with $\approx 100$ ms of physical time
for neural activity. In such a model the basic assumption is that the physical time required for
the transmission of information over synapses is somehow neglected in the psychological time.
The time ($\approx 100$ ms) required for the transmission of information
from retina to the inferiotemporal cortex (IT) through the primary visual cortex (V1) is mapped to a moment
of psychological time. It might be that by using 
$$
\Delta \approx 100 \rm{ms}$$ 
we shall get the right scale of the QL-coding.

However, it seems that the situation is essentially more complicated. There 
are experimental evidences
that the temporial structure of neural functioning is not homogeneous. 
The time required for completion of color
information in V4 ($\approx 60$ ms) is shorter that the time for the completion 
of shape analysis in IT 
($\approx 100$ ms). In particular it is predicted that there will be under 
certain conditions 
a rivalry between color  and form perception. This rivalry in time is one of 
manifestations of complex
level temporial structure of brain.  There may exist various pairs of scales inducing the 
QL-representations of information.

\section{The Hilbert Space Projection of Contextual Probabilistic
Dynamics}

Let us assume that the reference observables $a$ and $b$ evolve with
time: $x=x(t), \; \: y=y(t),$ where $x(t_0)=a$ and $y(t_0)=b.$ To simplify
considerations, we consider evolutions which do not change ranges of
values of the reference observables:
 $X_a=\{x_1,x_2\}$ and $ X_b=\{y_1,y_2\}$ do not depend on time. Thus,
 for any $t,$ $x(t)\in X_a$ and $y=y(t)\in X_b.$ 

In particular, we can consider the very special case when the dynamical 
reference observables correspond to classical stochastic processes: $x(t,\omega)), 
y(t,\omega)),$ where  $x(t_0, \omega)=a( \omega)$ and $y(t_0, \omega)=b( \omega).$  
Under the previous assumption these are random walks with two-points state 
spaces $X_a$ and $X_b.$ However, we recall that in general we do not assume the 
existence of Kolmogorov measure-theoretic representation.

Since our main aim is the contextual probabilistic realistic
reconstruction of QM, we should restrict our considerations to
evolutions with the trigonometric interference. We proceed under the
following assumption:

(CTRB) (Conservation of trigonometric behavior) \; {\it The set of
trigonometric contexts does not depend on time: ${\it
C}^{\rm{tr}}_{x(t)\vert y(t)} = {\it  C}^{\rm{tr}}_{x(t_0)\vert y(t_0)}=
{\it  C}^{\rm{tr}}_{a\vert b}.$}

By (CTRB) if a context $C \in {\it  C}^{\rm{tr}}_{x(t_0)\vert y(t_0)},$
i.e., at the initial instant of time the coefficients of statistical
disturbance $\vert \lambda(x(t_0)=x\vert y(t_0), C)\vert \leq 1,$ then
the coefficients $\lambda(x(t)=x\vert y(t), C)$ will always fluctuate in
the segment $[0,1].$ \footnote{Of course, there can be considered
more general dynamics in which the trigonometric probabilistic
behaviour can be transformed into the hyperbolic one and vice
versa.}

 For each instant of time $t,$ we can use QLRA, see (\ref{EX1}): a context $C$ can be
represented by a complex probability amplitude:
$$
\varphi(t, x)
\equiv \varphi_C^{x(t)\vert y(t)} (x)=\sqrt{p_C^{y(t)}(y_1) p^{x(t)\vert y(t)} (x\vert y_1)}
$$
$$
+ e^{i
\theta_C^{x(t)\vert y(t)}(x)} \sqrt{p_C^{y(t)} (y_2) p^{x(t)\vert y(t)}
(x\vert y_2)} .
$$
We remark that the observable $y(t)$ is represented by the 
self-adjoint operator $\hat y(t)$ defined by its
with eigenvectors:
\[e_{1t}^b= \left( \begin{array}{ll}
\sqrt{p_t( x_1\vert y_1)}\\
\sqrt{p_t( x_2\vert y_1)}
\end{array}
\right )
e_{2t}^b= e^{i \theta_C(t)} \left( \begin{array}{ll}
\sqrt{p_t( x_1\vert y_2)}\\
-\sqrt{p_t( x_2\vert y_1)}
\end{array}
\right )
\]
where $ p_t( x\vert y)=p^{x(t)\vert y(t)}(x\vert y),
\theta_C(t)=\theta_C^{x(t)\vert y(t)}(x_1) $ and where we set
$e_{jt}^b \equiv e_j^{y(t)}.$ We recall that $\theta_C^{x(t)\vert y(t)}
(x_2) = \theta_C^{x(t)\vert y(t)}(x_1) + \pi,$ since the matrix of
transition probabilities is assumed to be double stochastic for all
instances of time.

We shall describe dynamics of the wave function $\varphi(t, x)$
starting with following assumptions  (CP) and (CTP). Then these
assumptions will be completed by the set (a)-(b) of mathematical
assumptions which will imply the conventional Schr\"odinger
evolution.

(CP)(Conservation of $b$-probabilities)\;{\it The probability
distribution of the $b$-observable is preserved in process of
evolution:} $p_C^{y(t)}(y)=p_C^{b(t_0)}(y), y \in X_b,$ for any
context $C \in {\it  C}^{\rm{tr}}_{a(t_0)\vert b(t_0)}.$ This statistical
conservation of the $b$-quantity will have very important dynamical
consequences.  We also assume that the law of  conservation of
transition probabilities holds:

(CTP) (Conservation of transition probabilities)\;{\it Probabilities
$p_t( x\vert y)$ are conserved in the process of evolution:} $ p_t(
x\vert y)=p_{t_0}(x\vert y)\equiv p(x\vert y).$

Under the latter assumption we have:
$$
e_{1t}^b \equiv e_{1t_0}^b,\; 
e_{2t}^b=e^{i[\theta_C(t) - \theta_C(t_0)]} e_{2t_0}^b.
$$
For such an evolution of the $y(t)$-basis
$\hat{b}(t)=\hat{b}(t_0)=\hat{b}.$ Hence the whole stochastic
process $y(t,\omega)$ is represented by one fixed self-adjoint
operator $\hat{b}.$ This is a good illustration of incomplete QL
representation of information.

Thus under assumptions (CTRB), (CP) and (CTP) we have:
$$
\varphi(t)=u_1^b e_{1t}^b + u_2^b  e_{2t}^b=u_1^b e_{1t_0}^b + e^{i
\xi_C(t,t_0)} u_2^b e_{2t_0}^b ,
$$
where $u_j^b=\sqrt{p_C^{b(t_0)} (y_j)}, j=1,2,$ and $ \xi_C(t,t_0)=
\theta_C(t) - \theta_C(t_0).$ Let us consider the unitary operator
$\hat{U}(t, t_0): {\bf H} \to {\bf H}$ defined by this
transformation of basis: $e_{t_0}^b \to e_{t}^b.$ In the basis
$e_{t_0}^b=\{e_{1t_0}^b, e_{2t_0}^b\}$ the $\hat{U}(t, t_0)$ can be
represented by the matrix:
\[\hat{U}(t, t_0)= \left( \begin{array}{ll}
1&0\\
0 & e^{i\xi_C(t,t_0)}
\end{array}
\right ).
\]
We obtained the following dynamics in the Hilbert space ${\bf H}$:
\begin{equation} \label{E} \varphi(t)=\hat{U}(t, t_0) \varphi(t_0). \end{equation} This
dynamics looks very similar to the Schr\"odinger dynamics in the
Hilbert space. However, the dynamics (\ref {E}) is essentially more
general than Schr\"odinger's dynamics. In fact, the unitary operator
$\hat{U}(t, t_0)=\hat{U}(t, t_0, C)$ depends on the context $C,$
i.e., on the initial state $\varphi(t_0): \hat{U}(t, t_0) \equiv
\hat{U}(t, t_0, \varphi(t_0)).$ So, in fact, we derived the
following dynamical equation: $\varphi(t)=\hat{U}(t, t_0, \varphi_0)
\varphi_0,$ where, for any $\varphi_0, \hat{U}(t, t_0, \varphi_0)$
is a family of unitary operators.

The conditions (CTRB), (CP) and (CTP)  are  natural from the
physical viewpoint (if the $b$-observable is considered as an analog
of energy, see further considerations). But these conditions do not
imply that the Hilbert space image of the contextual  realistic
dynamics is a linear unitary dynamics. {\it In general the   Hilbert
space projection of the realistic prequantum dynamics is nonlinear.}
To obtain a linear dynamics, we should make the
following assumption:

(CI) (Context independence of the increment of the probabilistic phase)
{\it The $\xi_C(t,t_0)=\theta_C(t)-\theta_C(t_0)$ does not depend on $C.$}

Under this assumption the unitary operator $\hat{U}(t,
t_0)=\rm{diag}(1,e^{i \xi(t,t_0)}) $ does not depend on $C.$ Thus
the equation (\ref {E}) is the equation of the linear unitary
evolution. The linear unitary evolution (\ref{E})  is still
essentially more general than the conventional Schr\"odinger
dynamics. To obtain the Schr\"odinger evolution, we need  a few
standard mathematical assumptions:

(a). Dynamics is continuous: the map $(t,t_0) \to \hat{U}(t,t_0)$ is
continuous\footnote{We recall that there is considered the finite
dimensional case. Thus there is no problem of the choice of
topology.}; (b). Dynamics is deterministic; (c). Dynamics is
invariant with respect to time-shifts;  $\hat{U}(t, t_0)$ depends
only on  $t-t_0:$ $  \hat{U}(t,t_0)\equiv \hat{U}(t-t_0).$

The assumption of determinism can be described by the following
relation: $ \phi(t; t_0, \phi_0)=\phi(t; t_1, \phi(t_1; t_0,
\phi_0)) , \; t_0 \leq t_1 \leq t, $ where $\phi(t; t_0, \phi_0)=
\hat{U}(t,t_0) \phi_0.$

It is well known that under the assumptions (a), (b), (c)  the
family of (linear) unitary operators $\hat{U}(t, t_0)$ corresponds
to the one parametric group of unitary operators: $\hat{U}(t)=
e^{-\frac{i}{h} \hat{H}t},$ where $\hat{H}: {\bf H} \to {\bf H}$
is a self-adjoint operator. Here $h > 0$ is a scaling factor (e.g.,
the Planck constant). We have: $\hat{H} = \rm{diag}( 0, E),$ where 
$$
E= - h \Big[\frac{\theta_C(t)-\theta_C(t_0)}{t-t_0} \Big]. 
$$ 
Hence the Schr\"odinger evolution in the complex Hilbert space corresponds
to the contextual probabilistic dynamics with the linear evolution
of the probabilistic phase: 
$$
\theta_C(t) =\theta_C(t_0) -\frac{E}{h}(t-t_0).
$$ 

We, finally, study the very special case when the dynamical reference observables correspond to 
classical stochastic processes: $x(t,\omega)), y(t,\omega)).$ This is a special case of the V\"axj\"o model:
there exist the Kolmogorov representation of contexts and the reference observables.
Let us considxer a stochastic process (rescaling
of the process $y(t,\omega)):$ $ H(t, \omega) =0$ if  $y(t,
\omega)=y_1$ and $H(t,\omega)= E$  if $y(t, \omega)=y_2.$ Since the
probability distributions of the processes $y(t,\omega))$ and $H(t,
\omega)$ coincide, we have $p_C^{H(t)}(0)=p_C^{H(t_0)}(0);
p_C^{H(t)}(E)=p_C^{H(t_0)}(E).$

If $E > 0$ we can interpret $H(t,\omega)$ as the energy observable
and the operator $\hat{H}$ as its Hilbert space image. We emphasize
that the whole ``energy process'' $H(t,\omega)$ is represented by a
single self-adjoint nonnegative operator $\hat{H}$ in the Hilbert
space. This is again a good illustration of incomplete QL
representation of information.
This operator, ``quantum Hamiltonian'', is the Hilbert space
projection of the energy process which is defined on the
``prespace'' $\Omega$. In principle, random variables
$H(t_1,\omega), H(t_2,\omega), t_1 \not=t_2,$ can be very different
(as functions of $\omega).$ We have only the law of {\it statistical
conservation of energy:} $p_C^{H(t)}(z)\equiv p_C^{H(t_0)}(z), \;
z=0, E.$

\section{Concluding Remarks}

We created an algorithm -- QLRA -- for representation of a context (in fact, 
probabilitic data on this context) by a complex probability amplitude.

QLRA can be applied consciously by e.g. scientists to provide a consistent theory which is based on incomplete
statistical data. By our interpretation quantum physics works in this way. However, we can 
guess that a complex system could perform such a self-organization that QLRA would work 
automatically creating  in this system the two levels of organization: CL-level and QL-level.
Our hypothesis is that the brain is one of such systems with QLRA-functioning.
We emphasize that in the brain QLRA-representation is performed on the unconscious level
(by using Freud's terminology: in the unconsciousness). But the final result of application 
of QLRA is presented in the conscious domain in the form of feelings, associations and ideas, 
cf. \cite{KhrennikovS} and \cite{KhrennikovF}. We guess that some complex social systems are able to work in the QLRA-regime. As well
as for the brain for such a social system, the main advantage of working in the QLRA-regime 
is neglecting by huge amount of information (which is not considered as important). Since 
QLRA is based on choosing a fixed class of observables (for a brain or a social system 
they are {\it self-observables}), importance of information is determined by those
observables. The same brain or social system can use parallely a number of different QL 
representations based on applications of QLRA for different reference observables.

We can even speculate that physical systems can be (self-) organized through application of 
QLRA. As was already pointed out, Universe might be the greates user of QLRA.

{\bf Conclusion.} {\it The mathematical formalism of quantum mechanics can be applied outside
of physics, e.g., in cognitive, social, and political sciences, psychology, economics and 
finances.}


\begin{thebibliography}{99}

\bibitem{Khrennikov/P02}
A.~Yu. Khrennikov, editor, 
{\em Quantum Theory: Reconsideration of Foundations}. V\"axj\"o
Univ. Press, V\"axj\"o, 2002.

\bibitem{Khrennikov/P05}
G. Adenier and  A.Yu. Khrennikov, editors, 
{\em Foundations of Probability and Physics-3}. American Institute
of Physics, Melville, NY, {\bf 750},  2005.

\bibitem{Khrennikov/T}
A.~Yu. Khrennikov, To quantum mechanics through random fluctuations at the Planck time
  scale. {\em http://www.arxiv.org/abs/hep-th/0604011} (2006).


\bibitem{AR} D. Aerts and S. Aerts,   Applications of quantum statistics in psychological 
studies of decision-proceses. \textit{Foundations of
Science} \textbf{1}, 1-12 (1995).

\bibitem{AC} Accardi L {\it Urne e Camaleoni: Dialogo sulla realta, le leggi
del caso e la teoria quantistica},  Il Saggiatore, Rome, 1997 

\bibitem{Svozil}
K.~Svozil.
\newblock {\em Quantum Logic}.
\newblock Springer, Berlin, 1998.

\bibitem{Khrennikov/book:2004}
A.~Yu. Khrennikov, {\em Information Dynamics in Cognitive, Psychological and Anomalous
  Phenomena}. Kluwer Academic, Dordreht, 2004.


\bibitem{Choustova}
O.~Choustova, Quantum bohmian model for financial market.
{\em Physica A} {\bf 374}, 304--314 (2006).

\bibitem{Khrennikov/QLBrain}
A.~Yu. Khrennikov, Quantum-like brain: Interference of minds.
{\em BioSystems} {\bf 84}, 225--241 (2006).


\bibitem{Hameroff1} S. Hameroff,   Quantum coherence in microtubules.
A neural basis for emergent consciousness?  {\em  J. of Consciousness Studies}
{\bf 1} 91-118 (1994).

\bibitem{Hameroff2} S. Hameroff,    Quantum computing in brain microtubules? 
The Penrose-Hameroff Orch Or model
of consciousness. {\em  Phil. Tr. Royal Sc., London A} 1-28,  (1994).


\bibitem{Penrose1} R. Penrose, {\em The emperor's new mind.} Oxford Univ. Press, New-York,  1989.

\bibitem{Penrose2} R. Penrose,  {\em Shadows of the mind.} Oxford Univ. Press, Oxford, 1994.

\bibitem{Cohen} D. Cohen, {\it An inroduction to Hilbert space and quantum logic} (Springer, New York, 1989).

\bibitem{Foulis} D. J. Foulis, A half-century of quantum-logic. What have we learned? In:
{\it Quantum Structures and the Nature of Reality. Einstein meets Magritte,} {\bf 7}, pp. 1-36
(Kluwer, Dordrecht, 1990).

\bibitem{Svozil1} K. Svozil,  {\it Randomness and undecidability in physics} 
(World Sc. Publ., Singapore, 1994). 

\bibitem{Wright} R. Wright, Generalized urn models. {\it Foundations of physics}, {\bf 20}, 881-907 (1991). 


\bibitem{Neumann} J. von Neumann, {\it Mathematical foundations of quantum
mechanics.} Princeton Univ. Press, Princeton, N.J., 1955.

\bibitem{Birkhoff/vonNeumann}
G.~Birkhoff and J.~von Neumann, The logic of quantum mechanics.
{\em Ann. Math.} {\bf  37}, 823--643 (1936).

\bibitem{Khrennikov/Supp}
A.~Yu. Khrennikov, The principle of supplementarity: A contextual probabilistic
viewpoint to complementarity, the interference of probabilities, and the
incompatibility of variables in quantum mechanics.
{\em Foundations of Physics} {\bf 35}(10), 1655 -- 1693 (2005).


\bibitem{Conte} E. Conte, O. Todarello, A. Federici, F. Vitiello, M. Lopane, A.
Khrennikov and J. P. Zbilut, Some remarks on an experiment
suggesting quantum-like behavior of cognitive entities and
formulation of an abstract quantum mechanical formalism to describe
cognitive entity and its dynamics. {\it Chaos, Solitons  and
Fractals} {\bf 31},  1076-1088 (2006).


\bibitem{Mackey} G.~W. Mackey, {\em Mathematical Foundations of Quantum Mechanics.} 
W. A. Benjamin Inc., New York, 1963.

\bibitem{KhrennikovAN} A. Yu. Khrennikov, Classical and quantum mechanics on information spaces
with applications to cognitive, psychological,
social and anomalous phenomena.  {\em Foundations of Physics} {\bf 29}, 1065-1098 (1999).


\bibitem{Kolmogoroff}
A.~N. Kolmogoroff, {\em Grundbegriffe der Wahrscheinlichkeitsrechnung}.
Springer Verlag, Berlin, 1933.

\bibitem{Mori}
K.~Mori, On the relation between physical and psychological time.
In {\em Toward a Science of Consciousness}. Tucson
  University Press, Tucson, Arizona, 2002, p. 102.


\bibitem{Whitehead}
A.~N. Whitehead, {\em Process and Reality: An Essay in Cosmology}.
Macmillan Publishing Company, New York, 1929.

\bibitem{KhrennikovS}  A. Yu. Khrennikov,    Human subconscious as the $p$-adic
dynamical system. {\it J. of Theor. Biology} {\bf 193}, 179-196 (1998).

\bibitem{KhrennikovF} A. Yu. Khrennikov,  {\em Classical and quantum mental models and
Freud's theory of unconscious mind.} V\"axj\"o Univ.
Press, V\"axj\"o, 2002.


\end{thebibliography}
\end{document}